\documentclass[twocolumn,showpacs,preprintnumbers,amsmath,amssymb,superscriptaddress]{revtex4}
\usepackage{graphicx}
\usepackage{dcolumn,longtable}
\usepackage{bm}
\usepackage{epsfig}
\catcode`@=11
\def\lsim{\mathrel{\mathpalette\@versim<}}
\def\gsim{\mathrel{\mathpalette\@versim>}}
 \def\@versim#1#2{\lower0.2ex\vbox{\baselineskip\z@skip\lineskip\z@skip
       \lineskiplimit\z@\ialign{$\m@th#1\hfil##$\crcr#2\crcr\sim\crcr}}}
\catcode`@=12 
\def\beq{\begin{equation}}
\def\eeq{\end{equation}}

\def\tozero#1{\mathrel{\mathop{\sim}\limits_{\scriptscriptstyle
{#1\rightarrow0 }}}}
\begin{document}

\preprint{IFUM-855-FT}

\title{Re-evaluation of the  Gottfried sum using neural networks}

\author{Riccardo Abbate}%
\affiliation{Dipartimento di Fisica, Universit\`a di Milano}
\author{Stefano Forte}
\affiliation{Dipartimento di Fisica, Universit\`a di Milano}
\affiliation{INFN, Sezione di Milano, Via Celoria 16, I-20133 Milano, Italy}


\date{\today}
\begin{abstract}
We provide a determination of the Gottfried sum from all available
data, based on a neural network
parametrization of the nonsinglet structure function $F_2$. We find
$S_G=0.244\pm0.045$, closer to the quark model
expectation $S_G=\frac{1}{3}$ than previous results. We show
that the uncertainty from the small $x$ region is somewhat
underestimated in previous determinations.
\end{abstract}
\maketitle

The Gottfried sum
\beq
S_G(Q^2)\equiv\int_0^1\frac{dx}{x} \left[F_2^p(x,Q^2)-F_2^n(x,Q^2)\right]
\label{sgdef}
\eeq
provides a determination of the light flavor asymmetry of the nucleon
sea. The discovery by the NMC~\cite{nmca,nmcb,nmcc} that $S_G$ deviates from the simple quark
model expectation $S_G=\frac{1}{3}$ has provided first evidence for
an up-down asymmetry of the nucleon sea, a finding which has been
subsequently confirmed in different contexts, is routinely included in
modern parton fits, and has spawned a large theoretical
literature~\cite{gottpap,kumano}. Because the scale dependence
of the Gottfried sum is known up to next-to-next-to leading
order~\cite{kataeva,kataevb} its precise determination is potentially
interesting for tests of QCD and the determination of the strong coupling.

The experimental determination of a sum rule, 
and especially of the associated uncertainty, is
nontrivial because 
structure function data 
are only  available  at discrete values of $x$ and in general not all
given at the same $Q^2$. Therefore, one needs interpolation and
extrapolation in $x$ in order to cover the full range $0\le x\le 1$, 
and extrapolation in $Q^2$ in order to bring all data at the same
$Q^2$. Also, it is not obvious how to  combine data from different experiments without
losing information on experimental errors and correlations.

In Refs.~\cite{nnold,nnnew} the NNPDF collaboration has proposed 
a method for the parametrization of
structure functions and parton distributions, and has constructed 
a parametrization of the proton, deuteron,
and nonsinglet $F_2^{\rm NS}\equiv F_2^p-F_2^n$ structure functions
based on all available
experimental information, including experimental and theoretical
uncertainties and their correlation. This parametrization has been
recently used by various
authors~\cite{nnusers,nnusersa,nnusersb} as an
unbiased interpolation of existing data. 

\begin{figure}[b]
\begin{center}
\epsfig{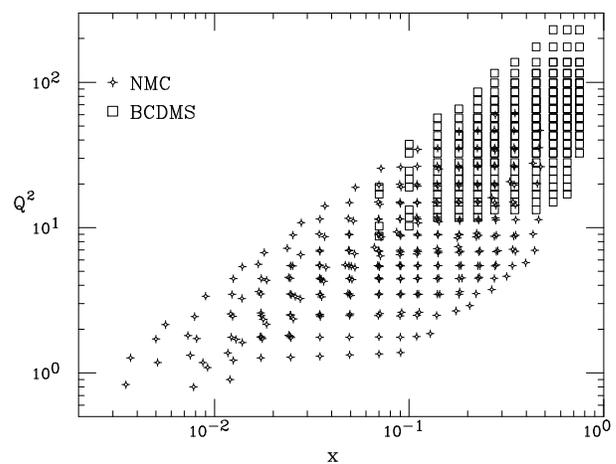}
\end{center}
\begin{center}
\caption{Data used to construct the neural network parametrization of
  $F_2^{\rm NS}$ of Ref.~\cite{nnold}.}
\label{fig:kinrange}
\end{center}
\end{figure}

Here, we  wish to provide a
determination of the Gottfried sum based on this parametrization,
which, in the nonsinglet case, relevant for the Gottfried sum,
is based on the structure function data from the  NMC~\cite{nmcf2} 
as well as  those from the BCDMS collaboration~\cite{bcdmsa,bcdmsb}, which are rather 
more precise and cover a different kinematic region (see
Fig.~\ref{fig:kinrange}). 

The parametrization of Refs.~\cite{nnold,nnnew} provides a Monte Carlo
sample of replicas of the
structure function for all $x$ and $Q^2$, so the Gottfried
sum and associated error 
can be straightforwardly determined by integrating over $x$  at fixed $Q^2$, and averaging over
the sample. The error on the parametrization blows up when 
extrapolating outside the measured region, so that the region where
reliable predictions are obtained can be inferred from the parametrization
itself.

First, we compare to the
result of Refs.~\cite{nmcb,nmcc}, where 
the contribution to the Gottfried integral
Eq.~(\ref{sgdef})  
from the measured region $0.004<x<0.8$ at $Q^2=4~\hbox{GeV}^2$ 
is determined to be
$S_G(0.004<x<0.8,4~\hbox{GeV}^2)=0.2281\pm0.0065~\hbox{ (stat.)}
\pm0.019~\hbox{ (syst.)}=0.2281\pm0.020$. The previous determination
of Ref.~\cite{nmcb}, based on about half the statistics, had $S_G(0.004<x<0.8,4~\hbox{GeV}^2)=0.221\pm0.008~\hbox{ (stat.)}
\pm0.019~\hbox{ (syst.)}=0.2281\pm0.021$.
Using the neural parametrization of Ref.~\cite{nnold} we get
\beq
S_G(0.004<x<0.8,4~\hbox{GeV}^2)=0.2281\pm 0.0437,
\label{sgval}
\eeq
where the error includes statistical and (correlated) systematic
uncertainties which are combined in the parametrization of
Ref.~\cite{nnold} (NNPDF result, henceforth). Despite the (accidental)
perfect agreement with NMC of
the central value of the NNPDF result Eq.~(\ref{sgval}), the
uncertainty is more than twice as large.
This is
suprizing, in view of the fact the data used in
neural parametrization include both NMC and BCDMS, and cover a
wider kinematic region.

In fact, the dataset  used by NMC in their
determination of the $F_2^p/F_2^d$ ratio which is used to compute the Gottfried
sum in Ref.~\cite{nmcc} is about
four times as large as that used by the same collaboration in their 
structure function determination of Ref.~\cite{nmcf2} which was used
to construct the neural network
parametrization~\cite{nnold}, 
essentially because the
detrmination of a structure function ratio allows more generous cuts
than the absolute determination of the structure function. However, it
is unclear whether this can explain the higher precision of the
determination  of Ref.~\cite{nmcc}, given that the error is dominated
by systematics, and the determination of $S_G$ requires
anyway knowledge of at least one structure function on top of the
structure function ratio.

\begin{table}[h]
\caption{The contribution to the Gottfried sum at $Q^2=4~\hbox{GeV}^2$
from the region $x_{\rm
    min}\le x\le 0.8$ as obtained
by NMC~\cite{nmcb} and
  with neural networks. The error is only statistical for NMC, while
  it is the total combined statistical and systematic uncertainty for
  NNPDF.  The total NMC systematics on $S_G(0.004\le x\le0.8)$ is
  equal to $0.019$.}
\begin{tabular}{|c||c|c||}
\hline
 $x_{\rm min}$&\multicolumn{2}{c|}{$S_G(x_{\rm
  min}<x<0.8$)} \\
\hline
 &NMC  &NNPDF \\
\hline\hline
$0.004$&$0.221 \pm 0.008 $&$0.2281 \pm 0.0437$ \\\hline
$0.010$&$0.213 \pm 0.005 $&$0.2378 \pm 0.0273$\\\hline
$0.020$&$0.203 \pm 0.004 $&$0.2334 \pm 0.0232$ \\\hline
$0.040$&$0.183 \pm 0.004 $&$0.2157 \pm 0.0217$\\\hline
$0.060$&$0.171 \pm 0.003 $&$0.1985 \pm 0.0202$\\\hline
$0.100$&$0.149 \pm 0.003 $&$0.1693 \pm 0.0169$\\\hline
$0.150$&$0.125 \pm 0.003 $&$0.1398 \pm 0.0133$\\\hline
$0.200$&$0.107 \pm 0.003 $&$0.1154 \pm 0.0107$\\\hline
$0.300$&$0.074 \pm 0.003 $&$0.0761 \pm 0.0074$\\\hline
$0.400$&$0.047 \pm 0.002 $&$0.0460 \pm 0.0052$\\\hline
$0.500$&$0.025 \pm 0.002 $&$0.0241 \pm 0.0035$ \\\hline
$0.600$&$0.012 \pm 0.002 $&$0.0102 \pm 0.0019$ \\\hline\hline
\end{tabular}
\label{tab:comp}
\end{table}
In order to understand this state of affairs, in Table~\ref{tab:comp}
we compare the contribution to the Gottfried integral
Eq.~(\ref{sgdef}) from the measured experimental region $x_{\rm
  min}\le x\le 0.8$~\cite{nmcb} with that obtained from neural
networks. The NNPDF determination uncertainty includes statistical and correlated systematic
errors, while the NMC experimental result Ref.~\cite{nmcb} only determines
the overall systematic uncertainty. Combining the total NMC
systematics (which is highly correlated between bins) with the statistical error of
Table~\ref{tab:comp}, the NNPDF and NMC total
uncertainties are seen to be in very good agreement up to the
next-to-smallest $x$ bin. However, when the smallest $x$ bin is
included the 
NNPDF uncertainty
almost doubles, while the 
NMC uncertainty (which is 
dominated by systematic) is essentially unchanged.

\begin{figure}[b]
\begin{center}
\epsfig{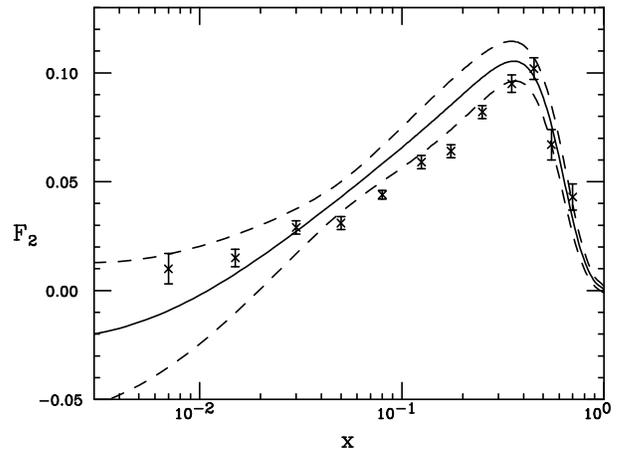}
\end{center}
\begin{center}
\caption{The nonsinglet structure function $F_2^{\rm NS}(x,Q^2)$ as a
  function of $x$ at $Q^2=4~\hbox{GeV}^2$. The solid line is the
  central value obtained from neural network and the dashed lines give
  the corresponding one-sigma error band (note errors are highly
  correlated betwen different values of $x$). The experimental points
  are from Ref.~\cite{nmcb}, the error bars are statistical only.}
\label{fig:fns}
\end{center}
\end{figure}
This suggests that the NMC uncertainty from the smallest $x$ bin might
be underestimated.
The reason for this is understood by inspection of Fig.~\ref{fig:fns},
where the NNPDF and  NMC determinations of the nonsinglet structure
function at $Q^2=4~\hbox{GeV}^2$ are compared. Note that the NMC error
bars are purely statistical, and that the NNPDF error band has high
point-to-point correlation (so the error on $S_G$ is much smaller than
the spread of the integrals of the one-sigma
curves). Note also that the NMC data points~\cite{nmcb} are obtained by combining
their determination of $F_2^d$ and of the ratio $F_2^p/F_2^d$, and
extrapolating the results at fixed $x$ to a common $Q^2$, while the
NNPDF results are obtained~\cite{nnold} by interpolating and
extrapolating the full set of NMC and BCDMS $F_2$ data: hence the two
determinations should agree within errors, but they are not expected to be on top
of each other. 

The two determinations are indeed seen to be in good
agreement. The agreement of the total
uncertainty on $S_G$ for $x\gsim 0.01$ proves that, once 
systematics is included, the total uncertainties also agree. At the smallest $x$
values, $x\lsim 0.01$, the uncertainty on $F_2$ blows up  nonlinearly
as a function of $x$, due
to the lack of smaller $x$ data which could constrain the
extrapolation. The NNPDF result, which is obtained integrating
$F_2^{\rm NS}$ reproduces this blowup. The NMC, based on summing over
bins (i.e. multiplying the value at the bin center by the bin width)
implicitly assumes that the error is linear across the bin and thus
underestimates the error on the last bin. 

We conclude that 
the NMC error on the Gottfried sum from the measured region is
smaller than the NNPDF error Eq.~(\ref{sgval}) entirely due to the
contribution of the smallest $x$ bin, and that this in turn is largely
due to the fact that the sum over bin by NMC underestimates the
nonlinear growth of the uncertainty at the edge of the data
region~\footnote{The final NMC determination~\cite{nmcc} is obtained
from a dataset which is about a factor 2 larger than that of
Ref.~\cite{nmcb}, but the bin-by-bin determination of $S_G$ shown in
Tab.~\ref{tab:comp} is not available for these data. However, the conclusion of the analysis of
table~\ref{tab:comp} and figure~\ref{fig:fns}, based on the data of Ref.~\cite{nmcb}
are unchanged because the uncertainty in the smallest $x$ bins is
largely dominated by systematics. Note also that at small $x$ the determination of
$F_2^{\rm NS}$  with the method of Refs.~\cite{nmca,nmcb,nmcc}, based
on the measurement of the $F_2^p/F_2^d$ ratio, is a priori more
precise than that based on the determination of their difference
because in the smallest $x$ bins there is a large cancellation so that
$FF_2^{\rm NS}$ is about one order of magnitude smaller than $F_2^p$
and $F_2^d$. Nevertheless, this is the case essentially for all $x$,
while the NNPDF and NMC errors only disagree in the smallest $x$ bin.}.

Let us now turn to the best determination of $S_G$
that can be obtained from neural networks. To this purpose, we note
that even though in principle the neural parametrization of $F_2$
provides a value for all $x$ and $Q^2$, when extrapolating outside the
data region this determination becomes unreliable: the uncertainty
grows rapidly, but eventually the uncertainty itself is
unreliable. Furthermore, whereas the neural nets do satisfy the
kinematic constraint $F2(1,Q^2)=0$, they do not satisfy the
theoretical constraint  $F2(0,Q^2)=0$, so the error on $S_G$ would
diverge if the sum rule were determined by simply integrating from $0$ to
$1$. This is as it should be, because the 
$x\to0$ region corresponds to the limit of infinite energy, and thus
it is even in principle experimentally unaccesible:  the associate
error  is therefore infinite unless one makes some theoretical assumption.

Therefore, we determine the Gottfried sum by integrating in $x$ at
fixed $Q^2$ for $x_{\rm min}\le x \le 1$, and adding to this integral a
contribution from the small $x$ region determined by extrapolation. Note that no
extrapolation is necessary in the large $x$ region, because the
coverage of the large $x$ region from the BCDMS data together with the
kinematical constraint at $x=1$ are sufficient to pin down the
structure function with good accuracy at large $x$~\cite{nnold}. 

The small $x$  extrapolation requires theoretical
assumptions. In Ref.~\cite{nmca}, it was assumed that
$F_2(x,Q^2)\tozero{x} A x^b$, and the constants $A$ and $b$ were
determined by fitting to the smallest $x$ data. However, the
assumption that the small--$x$ power behaviour has already set in in
the smallest measured $x$ bins does not seem justified. Indeed,
in
the singlet case
the small $x$ behaviour observed at HERA is not seen in the NMC data
and canot be predicted by them~\cite{nnnew,Dittmar:2005ed}. Also, on
theoretical grounds one would expect the asymptotic small $x$
behaviour to set in around $x\approx
10^{-3}$~\cite{Ermolaev:2005ny}. Hence, fitting the small $x$ exponent
to the data might lead to an underestimate of the uncertainty on the small
$x$ extrapolation, if the exponent $b$ comes out too large. 

\begin{table}[h]
\caption{Determination of the Gottfried sum with neural networks. The
 scale is given in GeV$^2$. The contribution from 
 $x< x_{\rm min}$ is obyained by extrapolation and given 100\%
 uncertainty (see text).
}
\begin{tabular}{|c|c|c|c|}
\hline
$Q^2$ & $x_{\rm min}$ &
$S_G(x_{\rm
  min}<x<1)$ & $S_G$ \\
\hline\hline
$1$ & $0.007$ & $0.2566\pm0.0773$  & $0.2849\pm 0.0917$ \\ \hline
$2$ & $0.005$& $0.2522\pm0.0389$   &$0.2548\pm0.0494$ \\ \hline
$3$ & $0.007$ &$0.2430\pm0.0299$  &$0.2479\pm 0.0454$ \\ \hline
$4$ & $0.008$ &$0.2380\pm0.0302$  &$0.2415\pm 0.0477$ \\ \hline
$5$ & $0.008$ & $0.2330\pm0.0340$ &$0.2329\pm0.0507$ \\ \hline
$10$ & $0.01$ &$0.2246\pm0.0428$  &$0.2278\pm0.0627$  \\ \hline
$30$ & $0.008$ & $0.2395\pm0.0632$ &$0.2450\pm0.0860$  \\ \hline\hline
$1.5 - 4.5$ & $0.006$  & $0.2438 \pm 0.0320$ & $0.2438 \pm 0.0449$\\ \hline
\end{tabular}
\label{tab:gottres}
\end{table}
Therefore, we extrapolate by assuming that the structure function  at small $x$
displays the behaviour  predicted by Regge theory~\cite{Ellis:1991qj}
  $F_2(x,Q^2)\tozero{x} A\sqrt{ x}$: even if in actual fact this
behaviour were
  to set in at smaller $x$, we would only be miscalculating the
  contribution to the integral from the matching region. Of course, we
  cannot exclude non--Regge behaviour at small $x$, but if the
  Gottfried integral Eq.~(\ref{sgdef}) exists at all, its integrand is
  unlikely to diverge much stronger than $\frac{1}{\sqrt{x}}$ at small
  $x$. Note that the small $x$ behaviour found by fitting to the data
  in Ref.~\cite{nmcb} is in fact somewhat softer than this, namely 
$F_2(x,Q^2)\tozero{x} A x^{0.59}$. 

Hence,  for all $x<x_{\rm min}$ 
we take $F_2(x,Q^2)=F^{sx}_2(x,Q^2)$, with
\beq
F^{sx}_2(x,Q^2)=A\sqrt{
    x}.
\label{sxf2}
\eeq
We fix the normalization
  coefficient $A$ by matching this behaviour to the neural network
  result at a somewhat larger value of $x$. This enables us to match
at a value
  of $x$ which is inside the data region, while only using
  $F^{sx}_2(x,Q^2)$ Eq.~(\ref{sxf2}) at the edge or outside  the data
  region itself. In practice, we match at $x=1.5 x_{\rm min}$; we have
  checked that results change very little if we move the
  matching point from $1.1x_{\rm min}$ to $2x_{\rm min}$. Matching to the
  neural network determination
of  $F_2(1.5x_{\rm min},Q^2)$ gives
 us a one-sigma error band $A=A_{\rm match}\pm\sigma_A$.

The  contribution to $S_G$ from $x<x_{\rm min}$ is determined as
the integral of
$F^{sx}_2(x,Q^2)$ computed with $A=A_{\rm match}$. This is given 100\%
uncertainty within the one-sigma error band of $A$, namely, the
extrapolation error
on is taken to be equal to the integral of
$F^{sx}_2(x,Q^2)$ computed with $A=|A_{\rm match}|+\sigma_A$.
The contribution to  $S_G$ is then added to the contribution from
the measured region, while the corresponding errors are added in
quadrature.

\begin{figure}[b]
\begin{center}
\epsfig{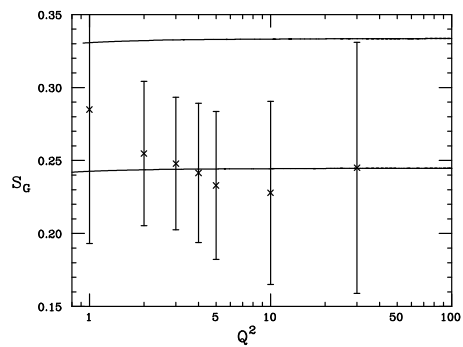}
\end{center}
\begin{center}
\caption{The value of the Gottfried sum $S_g$ (as given in
  table~\ref{tab:gottres}) as a function of the scale. The curve shows
  the scale dependence computed in perturbative QCD at
  NNLO~\cite{kataeva,kataevb}: upper curve assuming the quark model value
  $S_G(\infty)=\frac{1}{3}$,
lower curve assuming our best-fit $S_G(1\le Q^2\le 5)=.245$.}
\label{fig:sgplot}
\end{center}
\end{figure}
For each value of $Q^2$ we can thus find the value of $x_{\rm min}$
which minimizes the total uncertainty. There is a tradeoff in that if
$x_{\rm min}$ is raised, the error on the measured region decreases
rapidly, but there is an increase in 
size of the small $x$ extrapolation, which is 
100\% uncertain. The results for the Gottfried sum $S_G$
Eq.~(\ref{sgdef}), the contribution from the measured region, and the
value of $x_{\rm min}$ which minimizes the error are shown in
Table~\ref{tab:gottres}. 

In Fig.~\ref{fig:sgplot} these values are compared to the NNLO
perturbative prediction for the scale
dependence~\cite{kataeva,kataevb},~\footnote{Note that the numerical
  value of the NNLO $n_f$-independent contribution is given
  incorrectly due to typos in Ref.~\cite{kataeva}: the correct value is
  given in Ref.~\cite{kataevb} when $n_f=3$ and it is equal to
  $-0.850$ instead of $-0.809$ when $n_f=4$.} computed with the
assumption that as $Q^2\to\infty$ the naive quark model value
$S_G=\frac{1}{3}$ is reproduced.  This shows that, even though the
uncertainty in our determination is rather larger, and the central
value somewhat closer to the quark model prediction than the SMC
value~\cite{nmcc}, the quark model value and hence flavour symmetry of
the light quark sea are somewhat disfavoured. 

It is apparent from Fig.~\ref{fig:sgplot}
that the predicted perturbative dependence of $S_G$ is
very slight, and it is in fact entirely negligible on the scale of the
error on $S_G$. For example, the increase in $S_G$ from
$Q^2=1$~GeV$^2$ to $Q^2=10$~GeV$^2$ due to perturbative evolution is
less than 1\%. Hence, we may exploit the fact that neural networks
retain full information on correlations to combine the determination
of $S_G$ at different values of $Q^2$. When correlations are fully
taken into account, this can be done
by computing $S_G$
at an increasingly large number of values of $Q^2_{\rm min}\le Q^2\le
Q^2_{\rm max}$, until the result doesn't change as the number of
values of $Q^2$ at which $S_G$ is calculated in the given interval
does not change: because the correlation of $S_G(Q_1^2)$ and
$S_G(Q_2^2$ tends to one as $Q_1\to Q_2$, adding new points eventually
stops bringing in new information. 

However, because $S_G$ is very highly correlated between different
values of $Q^2$, and the uncertainty increases quite fast when the
scale is moved to low $Q^2$ (where data uncertainties are larger) or
high $Q^2$ (where there is little data coverage at small $x$), the
uncertainty on this averaged determination is only marginally smaller
than any of those which we obtained at fixed $Q^2$. Optimizing both
the $Q^2$ range and the choice of $x_{\rm min}$ we get our best value
\beq
S_G(1.5\le Q^2\le 4.5~\hbox{GeV}^2)=0.244\pm 0.045,
\label{sgbestval}
\eeq
which can be taken to hold for any $Q^2$ in the given range. The NNLO
$Q^2$ dependence of this result (assumed to hold at $Q^2=3$~GeV$^2$)
is also displayed in Fig.~\ref{fig:sgplot}.

A more precise determination of the Gottfried sum will only be
possible once more data will become available, such as those which
could be obtained injecting deuterons in HERA~\cite{Dittmar:2005ed},
from a high-energy upgrade of JLAB~\cite{Alekhin:2005at}, or from 
future facilities, such as the Electron-Ion collider~\cite{Deshpande:2002em} or a neutrino
factory~\cite{Mangano:2001mj}.

\bibliography{gott2}
\end{document}